\documentclass[aps,superscriptaddress,pra,secnumarabic,nobalancelastpage,amsmath,amssymb,nofootinbib,notitlepage]{revtex4-1}

\usepackage{graphics}      
\usepackage{longtable}     
\usepackage{url}           
\usepackage{bm}            
\usepackage{natbib}

\usepackage{amsmath}
\usepackage{amssymb}

\newcommand\bra[1]{\left \langle #1 \right|}
\newcommand\ket[1]{\left| #1 \right \rangle}

\renewcommand\Re[1]{\text{Re}\left(#1\right)}
\renewcommand\Im[1]{\text{Im}\left(#1\right)}

\begin{document}

\title{Matter-Wave Imaging of Quantum Density Fluctuations in
  Ultracold Bosons in an Optical Lattice}
\author{Scott N. Sanders}
\affiliation{Albert-Ludwigs-Universit\"{a}t Freiburg,
  Hermann-Herder-Str. 3, 79104 Freiburg, Germany}
\author{Florian Mintert}
\affiliation{Albert-Ludwigs-Universit\"{a}t Freiburg,
  Hermann-Herder-Str. 3, 79104 Freiburg, Germany}
\affiliation{Freiburg Institute for Advanced Studies, Albertstr. 19,
  79104 Freiburg, Germany}
\author{Andreas Buchleitner}
\affiliation{Albert-Ludwigs-Universit\"{a}t Freiburg,
  Hermann-Herder-Str. 3, 79104 Freiburg, Germany}
\date{\today}

\begin{abstract}
  We study the influence of quantum density fluctuations in ultracold
  atoms in an optical lattice on the scattering of matter waves. Such
  fluctuations are characteristic of the superfluid phase and vanish
  due to increased interactions in the Mott insulating phase.  We
  employ an analytical treatment of the scattering and demonstrate
  that the fluctuations lead to incoherent processes, which we propose
  to observe via decoherence of the fringes in a Mach-Zender
  interferometer. In this way we extract the purely coherent part of
  the scattering.  Further, we show that the quantum density
  fluctuations can also be observed directly in the differential
  angular scattering cross section for an atomic beam scattered from
  the atoms in a lattice. Here we find an explicit dependence of the
  scale of the inelastic scattering on the quantum density
  fluctuations.
\end{abstract}

\maketitle

\section{Introduction}

The atoms in a weakly interacting Bose Einstein condensate flow as a
superfluid over the crystalline potential landscape formed by a
periodic pattern of light intensity. Individual atoms in this system
can be spatially delocalized over the breadth of the optical lattice,
and, as a consequence, fluctuations in the number of atoms at
individual lattice sites occur even at zero temperature. As the
repulsion between the atoms grows, they are driven into a Mott
insulating phase, in which they maximally isolate themselves and
localize to a single lattice site~\cite{Fisher1989, Bloch2002,
  Bloch2008}. This rigid crystal, which reflects the behavior of
tightly bound electrons hopping in a solid, develops as the atoms
spontaneously lock to fixed, integer occupation of each lattice site,
and the fluctuations vanish.

Our objective is to study the influence of these quantum mechanical
fluctuations on the scattering of matter waves, and to illustrate the
susceptibility of the scattering to this purely quantum mechanical
effect, as well as the consequent impact of the transition from
non-interacting single particle physics to strongly interacting, many
body physics. Ultracold atoms in an optical lattice are an ideal
context in which to examine this behavior because they allow
controlled investigation of the transition between single and many
body physics, due to the experimental tunability of the physical
parameters of the crystal: the tunneling rate between lattice sites,
the interaction strength between atoms in the lattice and the geometry
of the lattice~\cite{Bloch2008, Yukalov2009, Greiner2009}.

It has been proposed that light scattering from atoms in an optical
lattice, situated within an optical cavity, will be sensitive to the
on-site number statistics~\cite{Ritsch2007}. Moreover, recent
experiments with high resolution optical systems have demonstrated
images resolving features on the scale of a single lattice site both
for atoms collected from a magneto-optic
trap~\cite{PhysRevLett.102.053001, Weiss2007} and for atoms in a
Bose-Einstein condensate~\cite{Greiner2009, Kuhr2010}. Due to the
interactions of the atoms with the light, however, only the parity of
the number of atoms at each site can be directly observed;
nonetheless, the fluctuations in the parity display a dependence on
the many body phase of the atoms in the lattice.

The scattering cross section of ultracold atoms in an optical lattice
seen by a matter wave probe is sensitive to the quantum many body
phase in the lattice~\cite{Sanders2010}. We seek to understand the
specific influence of the exotic properties of these materials on the
images produced by matter waves and to employ this knowledge to make
non-destructive measurements. Here, we study the effect of the purely
quantum mechanical density fluctuations at individual lattice sites
and show that these zero-temperature fluctuations can be directly
probed using a matter wave.

Fluctuations in the density distribution seen by an imaging matter
wave are expected to produce a dephasing effect and lead to incoherent
scattering. By examining the overlap of the matter wave scattered from
the atoms in an optical lattice with a coherent reference beam, in an
interferometric configuration, we will be able to explicitly see the
decohering effect of quantum density fluctuations on the interference
pattern, and to determine the size of the density fluctuations.
Subsequently, we will show that this incoherent scattering can even be
observed in the angular differential scattering cross section, and
that the scale of the fluctuations can be obtained directly by
examining the inelastic part of the scattering in an appropriate
regime of probe energy and lattice depth.

The matter wave probe is a free particle with mass $m$, initial wave
vector $\bm{k_0}$ and Hamiltonian $H_P = \bm{\hat{p}}^2/2m$, which
does not interact with the lattice light. The scattering target is
well modeled by the Bose Hubbard Hamiltonian, describing $N$ atoms
confined to the lowest band of a lattice with $N_L$ sites,
\begin{equation}
\label{eq:bh}
  \hat{H}_{\text{BH}} = -J \sum_{\left< \bm{R}, \bm{R}^{\prime} \right>}
  \hat{a}_{\bm{R}}^{\dagger} \hat{a}_{\bm{R^{\prime}}} + \frac{1}{2} U
  \sum_{\bm{R}} \hat{n}_{\bm{R}}(\hat{n}_{\bm{R}}-1).
\end{equation}
For a sufficiently large lattice, in which the effects of the edges of
the lattice are negligible, we may impose periodic boundary conditions
on both the atoms in the lattice and the transverse dimensions of the
probe. We may freely choose our coordinate system so that the $z$-axis is
aligned with $\bm{k_0}$. The initial probe wave function is then given by
\begin{equation}
\langle \bm{r} | \bm{k_0} \rangle = \frac{e^{i k_0 z}}{\sqrt{2\pi L^2}}, 
\end{equation}
where $k_0$ is the magnitude of $\bm{k_0}$, and $L$ is the length of
the lattice in the transverse dimensions.

As we wish our probe to weakly interact with the atoms in the lattice,
and to avoid interband excitations, we will insist on low energy
probes, for which s-wave scattering from the atoms in the target is
dominant, and we may treat the interaction between the probe and the
atoms in the lattice as a pseudopotential with scattering length,
$a_s$. The scattering interaction is then given by
\cite{Wodkiewicz1991},
\begin{equation}
\label{eq:V1}
\hat{V} = \sum_j
V_0 \, \delta(\hat{\bm{r}}-\hat{\bm{r}}_j),
\end{equation}
where the operators $\hat{\bm{r}}$ and $\hat{\bm{r}}_j$ give the
positions of the probe and the $j^{\text{th}}$ lattice atom,
respectively, and $V_0 = 2 \pi \hbar^2 a_s / m$. The full Hamiltonian
for the scattering interaction is $\hat{H} = \hat{H}_{\text{P}} +
\hat{H}_{\text{BH}} + \hat{V}$.

\section{Interference Measurement of Density Fluctuations}
\label{sec:interference}

\begin{figure}
\includegraphics{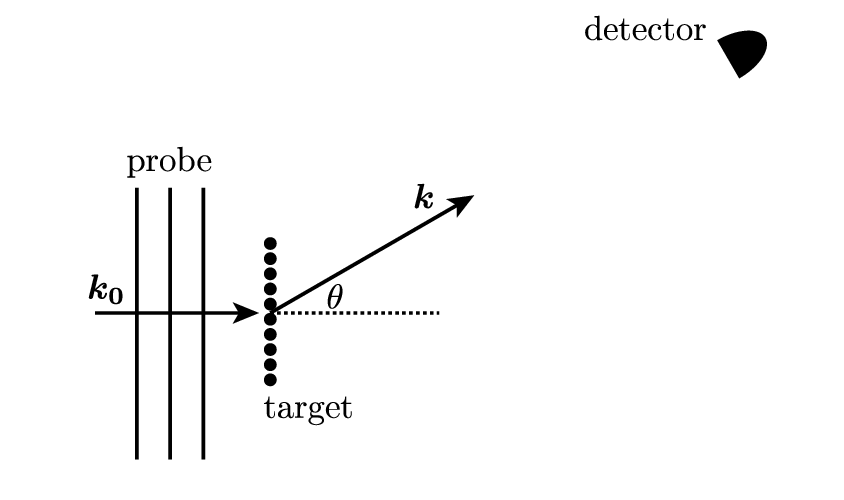}
\caption{\label{fig:standard-scattering}Scattering configuration in
  which the probe matter wave is incident from the left upon a
  one-dimensional lattice. A detector in the far-field measures the
  flux of probe atoms scattered in the direction $\theta$.}
\end{figure}

A standard scattering configuration, as depicted in
Fig.~\ref{fig:standard-scattering}, in which the angular differential
scattering cross section is measured, does not specifically
distinguish coherent from incoherent scattering.  The possibility that
disturbances are created in the target, such as long wavelength
phonons, exists despite the presence of evidence in the scattering
pattern for coherent scattering processes, such as Bragg
peaks. Insofar as these disturbances do not completely localize the
probe, interference of its wave function scattered from separated
points in the lattice can persist.

\begin{figure}
\includegraphics{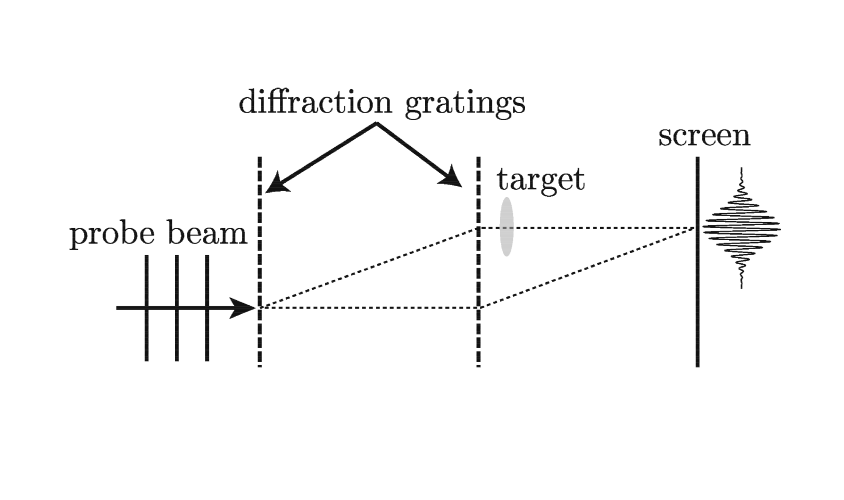}
\caption{\label{fig:mach-zender}Schematic of the Mach-Zender
  interferometer arrangement in which the probe beam is incident on a
  diffraction grating from the left. The $0^{\text{th}}$ and
  $1^{\text{st}}$ order diffracted beams form the lower and upper arms
  of the interferometer, respectively. The second diffraction grating
  causes the arms of the interferometer to overlap downstream at the
  position of the screen. A target is placed in the path of the upper
  arm only, so that only the coherent scattering will interfere with
  the lower, reference arm.}
\end{figure}

The situation is starkly different if the probe is coherently split
into two beams before scattering, so that there is a reference beam
that does not interact with the target. Interference between the atom
scattered from the target and the reference beam will then take place
only if the probe scatters without disturbing the target at all. We
will confirm the incoherent effect of the quantum density fluctuations
in the lattice on the scattered probe specifically by employing the
Mach-Zender configuration shown in
Fig.~\ref{fig:mach-zender}~\cite{Berman, Pritchard2009}.

In this arrangement, the upper arm scatters from the sample of cold
atoms in the optical lattice and then interferes downstream with the
lower arm. The interference pattern that forms there will be due
solely to the coherent part of the scattering, in which the state of
the atoms in the lattice is undisturbed~\cite{Walls1993,
  Sanders2009}. In practice, the narrow angular acceptance of the
region of overlap, where the detector is placed, also restricts
detection to the essentially forward scattered
atoms~\cite{Pritchard1995-1, Pritchard1995-2}. The forward scattered
beam may, however, acquire a phase shift and be attenuated due to
incoherent scattering. Both of these quantities, the phase shift and
attenuation, can be directly observed in the interference pattern that
forms. Here, we are particularly interested in analyzing the
attenuation, as this indicates the amount of incoherent scattering,
and, as we demonstrate below, depends on the zero-temperature density
fluctuations.

In order to determine this phase shift and attenuation, we consider
the S-matrix for the multi-particle system, which connects the initial
many body state of the probe and target with the asymptotic output
state, $\ket{\psi}$, long after the scattering interaction has
ceased~\cite{Watson1964-1},
\begin{equation}
\label{eq:S-basic}
\ket{\psi} = \hat{S} \ket{\bm{k_0}, n_0}. 
\end{equation}
The initial wave vector of the probe is labeled $\bm{k_0}$, and
$\ket{n_0}$ is the many body ground state of the $N$ atoms in the
lattice. The operator $\hat{S}$ can be expanded generically in terms
of the total energy $E$ of the $N+1$ atoms, the transition operator
$\hat{T}$ and the Hamiltonian $\hat{H}_0$, which excludes
interactions~\cite{Watson1964-2},
\begin{equation}
\hat{S} = 1 - 2 \pi i \, \delta \! \left( E- \hat{H}_0 \right) \hat{T}. 
\end{equation}
For the system we are discussing, the non-interacting Hamiltonian is
given by $\hat{H}_0 = \hat{H}_P + \hat{H}_{\text{BH}}$, and the
transition operator is given by~\cite{Sakurai1994-2},
\begin{equation}
\label{eq:T-def}
\hat{T} = \hat{V} + \lim_{\epsilon \rightarrow 0} \hat{V} \frac{1}{E_0 +
  E_{n_0} - \hat{H}_P - \hat{H}_{\text{BH}} + i \epsilon} \hat{T}. 
\end{equation}
The operator 
\begin{equation}
\hat{G}^{(N+1)}_0 = \left( E_0 + E_{n_0} - \hat{H}_P -
  \hat{H}_{\text{BH}} + i \epsilon \right)^{-1}
\end{equation}
is the free Green's operator for the many body system. $E_0$ is the
initial energy of the probe, and $E_{n_0}$ is the energy of the target
ground state. The state of the $N+1$ particle system that results from
the scattering, in terms of these operators, is
\begin{equation}
\ket{\psi} = \ket{\bm{k_0}, n_0} - 2 \pi i \, \delta \! \left( E-
  \hat{H}_0 \right) \hat{T} \ket{\bm{k_0}, n_0}.
\end{equation} 
The first term is the unscattered initial state. In the Mach-Zender
configuration, this term is coherent with the reference arm. The
second term involves both coherent and incoherent contributions. We
can facilitate extraction of the coherent contribution by expanding in
terms of eigenstates of the non-interacting probe and target. This is
accomplished by interposing a complete set of probe and target states
between the delta function and the operator $\hat{T}$. We may then
exclude contributions to the scattering that create disturbances in
the target or deflect the probe out of the forward direction. Choosing
the direction of $\bm{k_0}$ to align with the $z$-axis of our
coordinate system and recalling that the probe is subject to periodic
boundary conditions in the transverse dimensions, the expansion
yields
\begin{equation}
  \ket{\psi} = \ket{\bm{k_0}, n_0} - 2 \pi i \, \int \! dk_{z} \,
  \sum_{k_{x}, k_{y}} \sum_n \delta \left( E- \hat{H}_0 \right)
  \ket{k_{x}, k_{y}, k_{z}, n}  \bra{k_{x}, k_{y}, k_{z}, n} \hat{T}
  \ket{\bm{k_0}, n_0}. 
\end{equation}
Any scattering interaction which changes the state of the target, such
that $n \neq n_0$, would cause the probe to decohere completely, so
that we may exclude all such terms from the sum over target states. As
we mentioned above, the small angular acceptance of the detector
requires that we also exclude deflection of the probe in the $x-y$
plane. The remaining integral over $k_z$ may then be performed
immediately. The coherent part of the system after scattering is given
by
\begin{equation}
\label{eq:general-coherent-wave}
\ket{\psi_{coh}} = \left( 1- i \frac{2\pi m}{\hbar^2 k_0}
  \bra{k_{0}, n_0} \hat{T} \ket{k_0, n_0} \right) \ket{k_{0}, n_0},
\end{equation}
where, for simplicity, we write $\ket{k_0}$ as short hand for
$\ket{0,0,k_0}$, in which the components of the probe wave vector in
the $x$ and $y$ directions are
zero. Eq.~(\ref{eq:general-coherent-wave}) does not yet incorporate
the specific details of our target, and is written here in general for
any scattering matrix $\hat{T}$ and initial target state
$\ket{n_0}$. The diagonal $T$-matrix element of the initial state is
the crucial quantity that determines both the phase shift and the
contrast of the interference fringes. We may identify the complex
amplitude of the coherent wave, given in
Eq.~(\ref{eq:general-coherent-wave}), with the expansion of an
exponential when the strength of the interaction $V_0$ between the
probe and the atoms in the target is weak~\cite{Fermi1950}. The
coherent amplitude can then be separated into a part which corresponds
to a phase shift, $\phi = - \frac{2 \pi m}{\hbar^2 k_0} \Re{\langle
  \hat{T} \rangle}$, and an attenuation, $\xi = - \frac{2 \pi
  m}{\hbar^2 k_0} \Im{\langle \hat{T} \rangle}$, where both
expectation values are taken in the initial state, $\ket{k_0,
  n_0}$. The coherent part of the scattering in terms of these
quantities is
\begin{equation}
\ket{\psi_{\text{coh}}} = (1+i\phi -\xi) \ket{k_0, n_0}  \approx e^{i
  \phi - \xi} \ket{k_0, n_0}. 
\end{equation}
We may read off from this expression that the interference pattern
that forms will be shifted by $\phi$ and the fringes will have a
contrast $e^{-\xi}$~\cite{Walls1993, Sanders2009}.

It is appropriate to consider the case of weak interactions as we seek
to probe the target without disturbing its state. The values of $\phi$
and $\xi$ will then be dominated by the lowest-order terms in a Born
expansion of the operator $\hat{T}$, defined by
Eq.~(\ref{eq:T-def}). To first order in $V_0$, the transition operator
coincides with the interaction potential ($\hat{T} \approx \hat{V}$),
so that the $T$-matrix element becomes $\bra{k_{0}, n_0} \hat{V}
\ket{k_0, n_0}$. This quantity is purely real, and gives us the
leading order phase shift of the interference pattern. To first order
in $V_0$, however, there is no attenuation, and the scattering is
purely coherent. Loss of contrast in this system arises in the second
order correction, $\hat{T} \approx \hat{V} + \hat{V} \hat{G}^{(N+1)}_0
\hat{V}$. The second order contribution to the $T$-matrix element,
$\bra{k_{0}, n_0} \hat{V} \hat{G}^{(N+1)}_0 \hat{V} \ket{k_0, n_0}$,
is in general complex-valued. The real part of it will contribute a
small correction to the phase shift $\phi$, and the imaginary part
gives the leading order approximation of $\xi$.

In order to facilitate our intuition of the relationship between
$\phi$ and $\xi$ and the moments of the density distribution of the
atoms in the lattice, we will express the interaction potential in
second quantized form, in which it is given by
\begin{equation}
\label{eq:V}
\hat{V} = V_0 \, \hat{n}(\bm{\hat{r}_0}).
\end{equation}
Here $\hat{n}(\bm{r}) = \hat{\psi}^{\dagger}(\bm{r})
\hat{\psi}(\bm{r})$ is the density of the target at the position
$\bm{r}$. Inasmuch as the interaction potential depends linearly on
$\hat{n}(\bm{r})$, an expansion of the $T$-matrix in a Born series, in
orders of $V_0$, will contain terms that depend upon a corresponding
moment of the density distribution. For this reason, $\phi$, whose
leading contribution is linear in $V_0$, will depend solely on the
average density of the target. Likewise, $\xi$, which is second order
in $V_0$, will depend on the second moment of the density distribution
of atoms in the target, and therefore on the fluctuations.

Let us now explicitly evaluate the phase shift and decoherence of the
probe.  As we explained above, the phase shift requires that we
evaluate the diagonal $T$-matrix element in the first Born
approximation, which yields
\begin{equation}
\label{eq:first-order-T}
\bra{k_{0}, n_0} \hat{T}
  \ket{k_0, n_0} \approx \frac{V_0}{2\pi L^2} \int \!
  d^3r \, \bra{n_0} \hat{n}(\bm{r}) \ket{n_0}. 
\end{equation}
We find that, for any $\ket{n_0}$, to first order in the interaction
strength, in which only single scattering events are included in the
expansion of the $T$-matrix, the coherently scattered wave depends
only on the average density of the many body target,
\begin{equation}
\label{eq:coherent-wave-born}
  \ket{\psi_{coh}} = \exp \left(- i \frac{2 \pi a_s \rho}{k_0}
  \right) \ket{k_{0}, n_0}.  
\end{equation}
The approximation of the phase shift given by
Eq.~(\ref{eq:coherent-wave-born}) is proportional to the column
density, $\rho = N/L^2$, of the atoms in the target. This result is
not restricted to the case of cold atoms in an optical lattice, and is
known to determine, for example, the phase shift due to scattering
from a thermal gas~\cite{Pritchard1995-1, VanHove1954, Dalgarno1996,
  Sanders2009}.

By allowing terms that are second order in $V_0$, we enable scattering
channels in which the probe scatters from two different atoms within
the target, and thereby becomes sensitive to density fluctuations. The
coherent part of the scattered wave to second order is,
\begin{equation}
  \ket{\psi_{coh}} = \left( 1- \frac{2\pi i m}{\hbar^2 k_0}
    \left(\bra{k_{0}, n_0} \hat{V} \ket{k_0, n_0} + \bra{k_{0}, n_0} \hat{V}
      \hat{G}^{(N+1)}_0 \hat{V} \ket{k_0, n_0} \right) \right) \ket{k_{0}, n_0}. 
\end{equation}
As we argued above, the imaginary part of the second order term gives
us the exponent in the fringe contrast to leading order, so that we
must evaluate
\begin{equation}
\label{eq:contrast-exponent}
  \xi = -\frac{2\pi m}{\hbar^2 k_0} \, \text{Im} \left[ \bra{k_{0},
      n_0} \hat{V} \hat{G}^{(N+1)}_0 \hat{V} \ket{k_0, n_0} \right]. 
\end{equation}
We will isolate the dependence of $\xi$ on the density fluctuations
and determine the extent of their impact on the suppression of the
fringe contrast. This can be accomplished analytically in the regime
in which the energy of the probe, $E_0 = \hbar^2 k_0^2 /(2m)$, is
large compared to the bandwidth of the optical lattice, but not
sufficiently large to excite atoms out of the lowest band, so that we
continue to have single band dynamics.
The optical lattice potential along each spatial direction is $V_L(x)
= V_L \sin^2(x)$, where the depth of the lattice is conveniently
specified in units of the photon recoil energy $E_r = \hbar^2 k_L^2/(2
m_T)$. $k_L$ is the laser wave number, and $m_T$ is the mass of an
atom in the target. For a typical lattice depth of $V_L = 15 E_r$, the
width of the lowest band, determined by the energy difference between
the center and the edge of the first Brillouin zone, is found from the
exact solution for the optical lattice eigenvalue equation to be $0.03
E_r$. Likewise, the gap between the first and second bands, determined
by the energy difference between these bands at the edge of the first
Brillouin zone, is $6.28 E_r$. It is, therefore, readily possible to
employ a probe with energy much larger than the lowest band's width,
which is nonetheless insufficient to bridge the band gap.  In this
situation, we may neglect the loss of energy by the probe. The second
order term in the expansion of the diagonal $T$-matrix element that
appears in Eq.~(\ref{eq:general-coherent-wave}) is
\begin{equation}
\label{eq:second-order-T}
 \bra{k_{0}, n_0} \hat{V} \hat{G}_0^{(N+1)} \hat{V} \ket{k_0, n_0} =
 \frac{V_0^2}{2\pi L^2} \int \! d^3r \, d^3r^{\prime} \, e^{i \bm{k_0}
   \cdot (\bm{r^{\prime}}-\bm{r})} \bra{n_0} \hat{n}(\bm{r}) \,
 \hat{G}_0^{(N+1)}(\bm{r}, 
  \bm{r^{\prime}}; E_0+E_{n_0}) \, \hat{n}(\bm{r^{\prime}}) \ket{n_0}.
\end{equation}
The matrix element in this expression has a clear physical
interpretation. The probe wave function is scattered first at
$\bm{r^{\prime}}$, with a strength proportional to the density of
atoms in the target at this location. The scattered probe then evolves
freely according to the Green's function from $\bm{r^{\prime}}$ to
$\bm{r}$, where a second scattering event takes place. As we are
computing the coherently scattered projectile wave function, we only
include the contribution to this process which leaves the target
untouched, in the state $\ket{n_0}$. We then integrate over all such
two-point scattering paths.

Although the target must be left untouched, there may be transient
excitations between the two scattering events. Formally, this can be
seen by inserting a complete set of target states, $\sum_n \ket{n}
\bra{n}$, into the transition matrix element in
Eq.~(\ref{eq:second-order-T}). Note that the eigenstates of the
target, $\ket{n}$, are also eigenstates of the many body Green's
function,
\begin{equation}
\label{eq:many body-G-expansion}
\sum_n \hat{G}_0^{(N+1)}(\bm{r}, \bm{r^{\prime}}; E_0 + E_{n_0})
\ket{n}\bra{n} = \sum_n \bra{\bm{r}} \frac{1}{E_0 + (E_{n_0} - E_n) -
  \hat{H}_P + i\epsilon} \ket{\bm{r^{\prime}}} \, \ket{n}\bra{n}.  
\end{equation}
The coefficients in this expansion are the usual, single particle free
Green's functions at the shifted energy $\varepsilon_n = E_0 +
(E_{n_0}-E_n)$,
\begin{equation}
  \hat{G}_0(\bm{r}, \bm{r^{\prime}}; \varepsilon_n) =
  \bra{\bm{r}} \frac{1}{\varepsilon_n - \hat{H}_P + i \epsilon}
  \ket{\bm{r^{\prime}}}. 
\end{equation}
Physically, each term corresponds to an intermediate inelastic
scattering channel, in which the projectile has transferred energy
$E_n-E_{n_0}$ to the target. The largest possible energy transfer due
to scattering a single atom in the target within the lowest band is
the bandwidth. As we are considering the situation in which the
initial probe energy is large compared to the bandwidth of the lowest
band, the transient shift in its energy will be comparatively small,
so that $\varepsilon_n \approx E_0$. The single particle Green's
function that appears in the expansion in
Eq.~(\ref{eq:many body-G-expansion}) is, therefore, approximately
independent of the sum over target states. Inserting this expansion
into the second-order correction to the $T$-matrix element given in
Eq.~(\ref{eq:second-order-T}) yields
\begin{equation}
\label{eq:second-order-term-with-dens-dens-corr}
 \bra{k_{0}, n_0} \hat{V} \hat{G}_0^{(N+1)} \hat{V} \ket{k_0, n_0}
 \approx \frac{V_0^2}{2\pi L^2} \int \! d^3r \, d^3r^{\prime} \, e^{i
   \bm{k_0} \cdot (\bm{r^{\prime}}-\bm{r})} G_0(\bm{r},
 \bm{r^{\prime}}; E_0) \bra{n_0} \hat{n}(\bm{r}) \,
 \hat{n}(\bm{r^{\prime}}) \ket{n_0}. 
\end{equation}
In arriving at this expression, we have made use of the fact that the
projectile loses very little energy as it passes through the target,
and that interactions with individual atoms in the target are
predominantly s-wave scattering, due to the slow relative
velocities. The amplitude for each individual double scattering
sequence appropriately depends on the density-density correlator
between the two scattering points. 

As we seek the relationship between the on-site particle number
fluctuations in the lattice and the decay of the contrast of the
interference fringes, we must express the density-density correlator
in Eq.~(\ref{eq:second-order-term-with-dens-dens-corr}) in terms of
the on-site field operators. The field operator for the atoms in the
target can be expanded in terms of the Wannier function for the lowest
band of the lattice, $w(\bm{r})$, as
\begin{equation}
\hat{\psi}(\bm{r}) = \sum_{j=1}^{N_L} \hat{a}_j \, w(\bm{r}-\bm{R_j}), 
\end{equation}
where the operator $\hat{a}_j$ annihilates a particle at the
$j^{\text{th}}$ lattice site. The density $\hat{n}(\bm{r})$, in terms
of this expansion, is then given by
\begin{equation}
\hat{n}(\bm{r}) = \sum_{j,k=1}^{N_L} \hat{a}_j^{\dagger} \hat{a}_k \,
w^*(\bm{r}-\bm{R_j}) w(\bm{r}-\bm{R_k}).
\end{equation}
For the lattice depths that we are considering, $V_L \sim 15 E_r$, we
may employ the tight-binding approximation and neglect the
off-diagonal terms due to the small overlap between Wannier functions
centered on different lattice sites. The relationship between the
density at a position $\bm{r}$ and the density at a site, $j$, is then
approximately
\begin{equation}
\label{eq:napprox}
\hat{n}(\bm{r}) \approx \sum_{j=1}^{N_L} \hat{n}_j |w(\bm{r}-\bm{R_j})|^2.
\end{equation}
Substituting this into the density-density correlator will allow us to
explicitly extract the dependence on the fluctuations,
\begin{equation}
\bra{n_0} \hat{n}(\bm{r}) \,
\hat{n}(\bm{r^{\prime}}) \ket{n_0} \approx \sum_{j,k=1}^{N_L} \langle
\hat{n}_j \hat{n}_k \rangle |w(\bm{r}-\bm{R_j})|^2
|w(\bm{r^{\prime}}-\bm{R_k})|^2. 
\end{equation}
When the number of atoms $N$ in the target is much larger than 1, a
condensate of $N$ atoms in the ground state of the lattice is well
approximated by a coherent state with an average of $N$ atoms. In the
weakly interacting case, therefore, the target ground state is
appropriately described as a product of coherent states at each
lattice site with an average of $\bar{n}$ atoms per
site~\cite[p. 906]{Bloch2008}. In the strongly interacting case, the
target is a product of Fock states with $\bar{n}$ atoms per site. In
either case, we may simplify the cross correlators between different
lattice sites, $\langle \hat{n}_j \hat{n}_k \rangle = \langle
\hat{n}_j \rangle \langle \hat{n}_k \rangle$ ($j \neq k$), so that
these correlators are independent of the many body phase.  For a
uniform lattice, the average density does not vary from site to site,
so that $\langle n_j \rangle = \bar{n} = N/N_L$. The diagonal terms in
the sum yield a dependence on the second moment of the on-site number
distribution, which is also independent of the particular index,
$j$. The second moment can be re-expressed in terms of the on-site
fluctuations, $\sigma$, as $\sigma^2 = \langle \hat{n}_j^2 \rangle -
\bar{n}^2$. The contribution of the square of the average density may
be combined with the off-diagonal terms to give a sum over all pairs
of sites, leaving
\begin{equation}
\label{eq:dens-dens-with-fluctuations}
  \bra{n_0} \hat{n}(\bm{r}) \,
  \hat{n}(\bm{r^{\prime}}) \ket{n_0} \approx \sigma^2 \sum_{j=1}^{N_L}
  |w(\bm{r}-\bm{R_j})|^2 |w(\bm{r^{\prime}}-\bm{R_j})|^2 + \bar{n}^2
  \sum_{j,k=1}^{N_L} |w(\bm{r}-\bm{R_j})|^2
  |w(\bm{r^{\prime}}-\bm{R_k})|^2.  
\end{equation}
The second term in the density-density correlator, proportional only
to the average density, will not vary as the interaction strength is
increased across the transition between the superfluid and Mott
insulating phases; however, the first term, proportional to the
density fluctuations, will vanish as the interaction strength between
the atoms in the lattice increases, becoming zero in the Mott
insulator. The appearance of fluctuations in the diagonal $T$-matrix
element will lead to an additional loss of contrast in the
interference fringes.  Inserting this expression for the
density-density correlator into the expression for the second order
contribution to the $T$-matrix element given in
Eq.~(\ref{eq:second-order-term-with-dens-dens-corr}) yields
\begin{equation}
\label{eq:fluctuation-dens-dens}
 \bra{k_{0}, n_0} \hat{V} \hat{G}_0^{(N+1)} \hat{V} \ket{k_0, n_0} =
 \frac{V_0^2 \sigma^2 N_L}{2\pi L^2} F^{(2)}(\bm{k_0}) + A, 
\end{equation}
where $A$ refers to the contribution of the term proportional to
$\bar{n}^2$. This quantity is independent of the interaction strength
as the average density of atoms per lattice site is fixed for any
$U/J$ and will therefore not contribute to the variation of
the interference fringe contrast. The function $F^{(2)}(\bm{k_0})$
gives the amplitude for all possible two-point scattering events,
\begin{equation}
\label{eq:F2}
  F^{(2)}(\bm{k_0}) = \int \! d^3r \, d^3r^{\prime}
  \, e^{i \bm{k_0} \cdot (\bm{r^{\prime}}-\bm{r})} \left|w(\bm{r})
  \right|^2 G_0(\bm{r}, \bm{r^{\prime}}; E_0)  \left|
    w(\bm{r^{\prime}}) \right|^2. 
\end{equation} 
The imaginary part of $F^{(2)}(\bm{k_0})$ is calculated in
Appendix~\ref{sec:F2-derivation} for the concrete situation in which
the lattice is one-dimensional and oriented perpendicular to the
incident probe beam (see Fig.~\ref{fig:standard-scattering}). It is
then given by
\begin{equation}
  \Im{F^{(2)}(\bm{k_0})} = -\frac{m k_L}{\hbar^2 2^{3/2}
      \sqrt{\pi}} s^{1/4} \, \text{Erf} \left( \frac{k_0/k_L}{\sqrt{2}  
    s^{1/4}} \right),
\end{equation}
where $s = V_L/E_r$ is the depth of the lattice in units of the photon
recoil energy. Using this, we can explicitly write down the factor
that modifies the contrast of the interference fringes due to the
quantum mechanical density fluctuations in the lattice,
\begin{equation}
C \propto \exp \left(
-\frac{\rho a_s^2}{k_0/k_L} \frac{\sigma^2}{\bar{n}}
\frac{s^{1/4}}{\sqrt{\pi} 2^{3/2}} \, \text{Erf} \left(
  \frac{k_0/k_L}{\sqrt{2} s^{1/4}} \right)
 \right).
\end{equation}
Here we note that the functional dependence of the contrast on the
wave number of the probe and the depth of the lattice is in terms of
the dimensionless quantity $K = (k_0/k_L)/(\sqrt{2}
s^{1/4})$. Employing $K$, the factor giving the reduction of the
contrast due to the fluctuations in the lattice takes the form
\begin{equation}
\label{eq:contrast}
C \propto \exp \left(
-\frac{\rho a_s^2}{4 \sqrt{\pi}} \frac{\sigma^2}{\bar{n}}
\frac{1}{K} \, \text{Erf} \left( K \right)
 \right).
\end{equation}
For a typical lattice depth, $s=15$, and for the range of probe wave
numbers which avoid interband excitations at this lattice depth,
$k_0/k_L \leq \sqrt{6.28}$, the variation of the exponent with $K$ is
minor, and the function, $K^{-1} \, \text{Erf}(K)$ is of order
$1$. The remaining dependence of the contrast is such that it decays
exponentially with increasing fluctuations in the on-site atom number
and with increasing column density of the sample.

The attenuation of contrast given in Eq.~(\ref{eq:contrast}) can be
directly obtained from experimental measurement of the interference
fringes as the ratio of the contrast of the fringes at arbitrary
values of the parameter $U/J$ to the contrast deep in the Mott
insulator regime, where $\sigma^2 \rightarrow 0$. Although it is
common to alter the interaction strength $U/J$ by adjusting the depth
of the lattice, here we consider adjusting $U/J$ by manipulating the
scattering length of the lattice atom collisions through a Feshbach
resonance, so that we retain a constant lattice depth, and $K$ remains
constant.

We have shown that the quantum mechanical density fluctuations at
individual lattice sites have a decohering effect on a matter wave
passing through a Mach-Zender interferometer, and that the contrast in
the interference fringes is exponentially suppressed by increasing
fluctuations. It was necessary to employ an interferometric
arrangement to explicitly extract the coherent component of the
scattering, and in this way, we have definitively verified the
enhancement of incoherent scattering due to these density
fluctuations. In the following section, we will exploit these results
to identify the influence of the density fluctuations on the
scattering cross section and demonstrate the conditions under which
the fluctuations are directly observable therein.

\section{Density Fluctuations and Inelastic Scattering}
\label{sec:inel-scatt}
We begin our analysis of the differential scattering cross section
with a general result for scattering from a many body system in the
first Born approximation. The target is initially in the ground state,
$\ket{n_0}$. The cross section is then given by~\cite{Sanders2010,
  VanHove1954}
\begin{equation}
\label{eq:angdiffcs}
  \frac{d\sigma}{d\Omega} = a_s^2 \sum_n \sqrt{1 -
    \frac{E_n-E_{n_0}}{\hbar^2 k_o^2/(2m)}} \ 
  \left| \bra{n} \sum_{j=1}^{N} e^{i \bm{\kappa}
      \cdot \hat{\bm{r}}_j} \ket{n_o} \right|^2,
\end{equation}
where $\bm{\kappa} = \bm{k_0}-\bm{k}$ is the difference between the
incident probe wave vector $\bm{k_0}$ and the outgoing probe wave
vector $\bm{k}$. The magnitude of $\bm{k}$ is potentially reduced from
the incident probe wave number, due to a transfer of energy from the
probe to the lattice, and is determined by $\hbar^2 k^2/(2m) = \hbar^2
k_0^2/(2m) + (E_{n_0}-E_n)$. The outgoing probe wave vector points in
the direction $\theta$ of the detector (see
Fig.~\ref{fig:standard-scattering}).

Bragg peaks, which correspond to a coherent sum of waves scattered
from individual atoms in the lattice, appear in the elastic scattering
cross section~\cite{Sanders2010}. Additionally, we showed in
Sec.~\ref{sec:interference} that increased density fluctuations
correspond to increased incoherent scattering of the probe. This
suggests that the on-site density fluctuations will influence the
scattering cross section through inelastic channels, in which the
probe delivers energy to the atoms in the lattice. This intuition
motivates us to examine the cross section under the same energetic
conditions as in our analysis of the interference pattern in
Sec.~\ref{sec:interference}, described immediately following
Eq.~(\ref{eq:contrast-exponent}). There, the contribution of single
band inelastic channels was emphasized by choosing an initial probe
energy sufficient to allow excitation to all of the modes of the
lowest band of the lattice, but which did not exceed the splitting
between the lowest band and the first excited band.

The transition amplitude between the ground state, $n_0$, and an
excited state, $n$, due to a momentum boost, which is given by the
matrix element in the expression for the cross section in
Eq.~(\ref{eq:angdiffcs}), is non-zero in the first Born approximation
only for final target states in which at most a single atom in the
target is displaced out of the ground state of the lattice. The energy
of the final target modes that we must consider, $E_n$, differ,
therefore, from $E_0$ by the energy of a single particle
excitation. Thus, the conditions are viable to enforce the condition
$E_n-E_{n_0} \ll \hbar^2 k_o^2/(2m)$ for all accessible final target
states. At such high probe energies, the factor under the square root
in Eq.~(\ref{eq:angdiffcs}), which weights the contribution of the
scattering into the final target mode, $n$, is nearly uniform and
approximately unity. This factor appears also in $\bm{\kappa}$;
however, in the high-energy approximation, we take $\bm{\kappa}$ to be
independent of the energy transfer. This is equivalent to the
so-called static approximation~\cite{VanHove1954}. With these
simplifications, the cross section becomes
\begin{equation}
  \frac{d\sigma}{d\Omega} = a_s^2  
\bra{n_0} \sum_{j=1}^{N} e^{-i \bm{\kappa}
      \cdot \hat{\bm{r}}_j} \left( \sum_n \ket{n}\bra{n} \right) \sum_{k=1}^{N}
    e^{i \bm{\kappa} \cdot \hat{\bm{r}}_k} \ket{n_o}.
\end{equation}
In the subspace of the lowest band of the lattice, the sum over
projections onto individual target modes, $n$, is an identity and can
be removed. The cross section can then be expressed in terms of the
density-density correlator, given in
Eq.~(\ref{eq:dens-dens-with-fluctuations}), by rewriting it in second
quantized notation. This is accomplished via the identification of
$\sum_{k=1}^{N} e^{i \bm{\kappa} \cdot \hat{\bm{r}}_k}$ with $\int \!
d^3r \, e^{i \bm{\kappa} \cdot \bm{r}} \hat{n}(\bm{r})$, which leads
to the expression for the cross section,
\begin{equation}
  \frac{d\sigma}{d\Omega} = a_s^2  
\int \! d^3r d^3r^{\prime} \,  e^{-i \bm{\kappa}
      \cdot (\bm{r}-\bm{r^{\prime}})}  \bra{n_0} \hat{n}(\bm{r})
    \hat{n}(\bm{r^{\prime}}) \ket{n_o}.
\end{equation}

\begin{figure}
\includegraphics{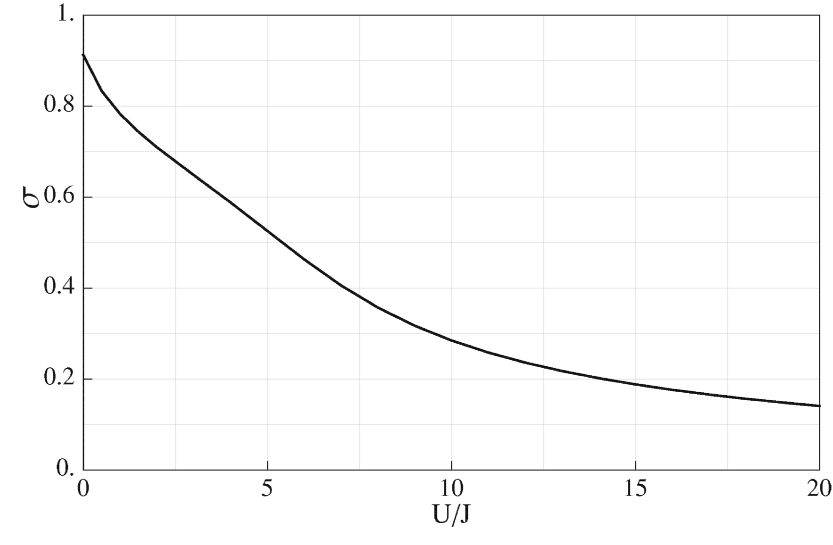}
\caption{\label{fig:fluctuations} The on-site number fluctuations,
  $\sigma$, plotted as a function of the parameter $U/J$, for a
  lattice with $N=6$ atoms and $N_L = 6$ lattice sites and a depth of
  $s=15$ times the photon recoil energy.}
\end{figure}

\begin{figure}
\includegraphics{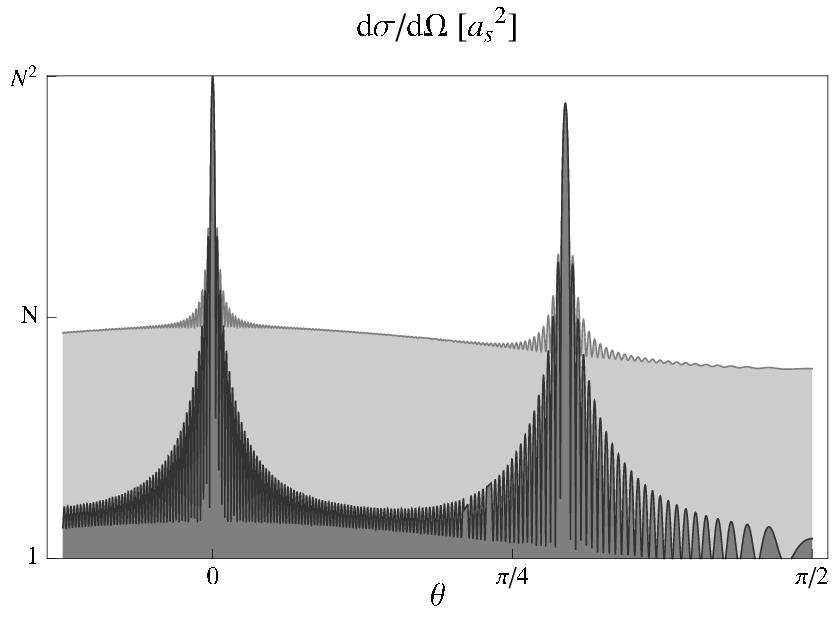}
\caption{\label{fig:cross-section}The cross section given by
  Eq.~(\ref{eq:final-cs}) for scattering a matter wave from the atoms
  in a one-dimensional lattice, oriented perpendicular to the incident
  probe wave vector, as a function of the angle $\theta$ from the
  forward direction. The first term in the cross section, which gives
  Bragg peaks, are computed for 100 atoms in 100 lattice sites. The
  second term, giving the inelastic background, is plotted using the
  values of $\sigma$ given in Fig.~\ref{fig:fluctuations} and a
  tunneling matrix element $J=0.0047 E_r$.  The plot in light gray
  corresponds to $U/J = 0$, in which the target is entirely
  superfluid. The small deviation from $N$ is a result of using the
  exact, non-interacting ground state, rather than the coherent state
  approximation. The cross section for scattering from a target with
  $U/J = 20$ is shown in dark gray.}
\end{figure}

We may expand the density $\hat{n}(\bm{r})$ according to
Eq.~(\ref{eq:napprox}), in terms of on-site densities and the Wannier
function of the lowest band. When the number of lattice sites is
sufficiently large that the effect of the edges may be neglected, the
cross section separates into two factors. One of which is a smooth
background determined by the shape of the Wannier function, which
reflects the depth of the lattice. The other depends only on the
distribution of atoms in the lattice. Thus for $N_L \gg 1$, we find
\begin{equation}
\label{eq:cs-with-site-correlators}
\frac{1}{a_s^2} \frac{d\sigma}{d\Omega} \approx \left(
  \sum_{j=1}^{N_L} \langle \hat{n}_j^2 \rangle + \sum_{j \neq k}
  \langle \hat{n}_j \hat{n}_k \rangle e^{i \bm{\kappa} \cdot
    (\bm{R_k} -\bm{R}_j)} \right) \left| \int \! d^3r \, e^{-i
      \bm{\kappa} \cdot \bm{r}} | w(\bm{r}) |^2 \right|^2. 
\end{equation}
A completely filled lattice, i.e., $N \geq N_L$, with a large number
of lattice sites will have a correspondingly large number of atoms, so
that the superfluid ground state is well-described by a coherent state
with an average of $N$ atoms. Then we have as before that $\langle
\hat{n}_j \hat{n}_k \rangle = \bar{n}^2$ ($j\neq k$) and $\langle
\hat{n}_j^2 \rangle = \sigma^2 + \bar{n}^2$.  The expression for the
cross section, valid in both the superfluid and Mott insulating
phases, is given by,
\begin{equation}
\label{eq:final-cs}
\frac{1}{a_s^2} \frac{d\sigma}{d\Omega} = \left( \bar{n}^2 \left|
    \sum_{j=1}^{N_L} e^{-i \bm{\kappa} \cdot \bm{R}_j} \right|^2 +
  \sigma^2 N_L \right) \left| \int \! d^3r \, e^{-i \bm{\kappa} \cdot
    \bm{r}} |w(\bm{r})|^2 \right|^2. 
\end{equation}
In this expression, the first term gives elastic Bragg peaks that
scale as the square of the number of atoms in the lattice. The
remaining term gives the inelastic cross section, which contributes a
smooth background. In the high probe energy regime, the inelastic
background is directly proportional to the fluctuations. Its amplitude
will therefore be reduced as the interaction strength between the
atoms in the lattice is increased relative to the tunneling matrix
element. The fluctuations are shown as a function of the parameter
$U/J$ in Fig.~\ref{fig:fluctuations} for a lattice with $N=6$ atoms
and $N_L=6$ lattice sites. This is sufficient due to the rapid
convergence of the shape of the decay of the fluctuations with
increasing lattice size~\cite{Dong2010}.  The sum over exponentials in
the first term yields Bragg peaks located at positions for which
$\bm{\kappa}$ is a reciprocal lattice vector. These arise due to the
coherent overlap of the scattered waves from individual atoms in the
lattice.
Our result for the cross section recreates the expected behavior, in
which, deep in the Mott insulator phase, with $\sigma^2 \rightarrow
0$, no inelastic background remains, whereas the superfluid phase, in
which $\sigma^2 \approx \bar{n}$, yields an inelastic background
proportional to the number of atoms in the lattice, $N$ (see
Fig.~\ref{fig:cross-section}).

The final factor, which is the Fourier transform of the Wannier
function, leads to an overall envelope with an approximately Gaussian
shape. In the region between the extremes of very weakly or very
strongly interacting atoms in the lattice, the fluctuations can be
directly observed by measuring the scale of the inelastic
background. In particular, the difference between the cross section
for an arbitrary value of $U/J$ and the cross section deep in the Mott
insulator regime will eliminate the elastic Bragg peaks, and leave
only a Gaussian background that scales as $\sigma^2 N_L$. This result
is intuitively satisfying in that we can trace the impact of the
density fluctuations to the incoherent part of the scattering. This is
made explicit through our analysis of the interferometer, and likewise
is consistent with our expectations for the scattering cross section,
in which we associate the inelastic part of the cross section with
incoherent scattering processes.

\section{Conclusions}

We have shown that matter wave scattering is susceptible to the purely
quantum mechanical fluctuations in on-site atom number in an optical
lattice. The fluctuations lead to incoherent scattering of the probe
atom, which can be quantified and observed by the loss of contrast of
the interference fringes in a Mach-Zender interferometer. The fringes
form due to the interference between the scattered matter wave and the
unscattered, reference arm of the interferometer. To first order in
the interaction strength, $V_0$, the interference pattern is shifted
due to a quantum mechanical phase proportional to the density of atoms
in the lattice. Decoherence occurs in the interferometer as a second
order effect. The second order term in a Born series expansion of the
many body scattering $T$-matrix, proportional to $V_0^2$, corresponds
physically to double scattering of the probe within the cold atom
sample. When this multiple scattering is taken into account, the
scattered matter wave becomes sensitive to the local density
fluctuations in the target, which cause decoherence of the probe atom.

In the regime in which the scattered probe atom has an energy that is
large compared to the bandwidth of the lowest band of the optical
lattice, the fluctuations can be directly obtained from the amplitude
of the inelastic scattering background. Whereas the elastic part of
the scattering cross section corresponds to coherent scattering, the
inelastic part contains the incoherent contribution, in which atoms in
the lattice are excited out of the ground state. This incoherent
process exhibits a simple and direct dependence on the density
fluctuations, providing a method to directly observe them.

\begin{acknowledgements}
The authors gratefully acknowledge useful discussions with Eric Heller. 
\end{acknowledgements}

\appendix

\section{Derivation of $F^{(2)}(\bm{k_0})$}
\label{sec:F2-derivation}

We will evaluate the two-point scattering function given in
Eq.~(\ref{eq:F2}),
\begin{equation}
  F^{(2)}(\bm{k_0}) = \int \! d^3r \, d^3r^{\prime}
  \, e^{i \bm{k_0} \cdot (\bm{r^{\prime}}-\bm{r})} \left|w(\bm{r})
  \right|^2 G_0(\bm{r}, \bm{r^{\prime}}; E_0)  \left|
    w(\bm{r^{\prime}}) \right|^2,
\end{equation} 
explicitly for the case of a one-dimensional lattice, in which the
atoms are tightly confined in the $y$ and $z$-directions, and the wave
vector $\bm{k_0}$ of the incident probe atom points along the
$z$-direction. Strictly speaking, the range of the integration is over
the length of the lattice; however, the locality of the Wannier
function allows us to extend the integration to all space. For a
harmonic confinement that is deep in the $y$ and $z$-directions, the
Wannier function approaches $|w(\bm{r})|^2 \approx |w(x)|^2 \,
\delta(y) \, \delta(z)$. The one-dimensional case, with the
substitution $u=x-x^{\prime}$, therefore yields,
\begin{equation}
\label{eq:F2-1D}
F^{(2)}(\bm{k_0}) = -\frac{m}{2\pi \hbar^2} \int_{-\infty}^{+\infty}
\! du \, \frac{e^{i k_0 |u|}}{|u|} \int_{-\infty}^{+\infty} \!
dx^{\prime} \, |w(u+x^{\prime})|^2 |w(x^{\prime})|^2. 
\end{equation}
We can use the harmonic approximation for the Wannier function in a
lattice of depth $s=V_0/E_r$,
\begin{equation}
w(x) \approx \left( \frac{2m_T}{\hbar^2 \pi} E_r \sqrt{s} \right)^{1/4}
\exp \left(-  \frac{2m_T}{\hbar^2} E_r \sqrt{s} x^2/2 \right).
\end{equation}
This allows us to compute the integral over $x^{\prime}$ in
Eq.~(\ref{eq:F2-1D}) analytically. The imaginary part of
$F^{(2)}(\bm{k_0})$ is then given by
\begin{equation}
\Im{F^{(2)}(\bm{k_0})} = -\frac{m \, m_T^{1/2} (s
  E_r^2)^{1/4}}{2\pi^{3/2} \hbar^3} 2 \int_0^{\infty} \! du \,
\frac{\sin(k_0 u)}{u} e^{-\frac{m_T}{\hbar^2} \sqrt{s} E_r u^2}. 
\end{equation}
Notice that we have taken advantage of the evenness of the integrand
to reduce the range of the integration to $[0,\infty)$ and thereby
eliminated the need for the absolute value operation on $u$.
The remaining integral over $u$ is then given by
\begin{equation}
  \Im{F^{(2)}(\bm{k_0})} = -\frac{m k_L}{\hbar^2 2^{3/2} \sqrt{\pi}}
  s^{1/4} \text{Erf}\left(\frac{k_0/k_L}{\sqrt{2} s^{1/4}}\right).
\end{equation}

\bibliography{Catalog}

\end{document}